\documentclass[12pt]{article}

\usepackage{setspace}
\newcommand{\beq}{\begin{equation}}
\newcommand{\eeq}[1]{\label{#1}\end{equation}}
\newcommand{\bea}{\begin{eqnarray}}
\newcommand{\eea}[1]{\label{#1}\end{eqnarray}}
\newcommand{\bh}{\stackrel{bh}{=}}
\begin{document}
\setlength{\topmargin}{-1cm} \setlength{\oddsidemargin}{0cm}
\setlength{\evensidemargin}{0cm}
\begin{titlepage}
\begin{center}
{\Large   Topologically Massive Gravity at the Chiral Point is Not Unitary}

\vspace{20pt}

{\large G. Giribet$^a$, M. Kleban$^b$ and  M. Porrati$^b$}

\vspace{12pt}

$a$) Department of Physics, Universidad de Buenos Aires and CONICET\\
Ciudad Universitaria, Pabell\'on I\\ 1428, Buenos Aires, Argentina

\vspace{12pt}

$b$) Center for Cosmology and Particle Physics\\
Department of Physics\\ New York University\\
4 Washington Pl.\\ New York, NY 10003, USA

\end{center}
\vspace{20pt}

\begin{abstract}
A recent paper~\cite{SLS} claims that topologically massive
gravity contains only chiral boundary excitations at a particular
value of the Chern-Simons coupling. On the other hand, propagating
bulk degrees of freedom with negative norm were found even at the chiral point
in~\cite{DCWW}. The two references use very different methods,
making comparison of their respective claims difficult. In this
letter, we use the method of~\cite{SLS} to construct a tower of physical
propagating bulk states satisfying standard AdS boundary
conditions.  Our states have finite norm, with sign opposite to that of right-moving 
boundary excitations. Our results thus agree with~\cite{DCWW} and disagree
with~\cite{SLS}.

\end{abstract}

\end{titlepage}

\newpage

Recently, interest in pure AdS gravity has been revived following
Witten's work~\cite{W}, which seemed to offer a chance of finding an exact
solution to {\em a} quantum gravity (albeit a particularly simple one).
Shortly afterward~\cite{SLS,SLS2} appeared, which argued for the existence of a theory even simpler than pure
gravity. The claim was that in
topologically massive gravity (TMG)~\cite{DJT1,DJT2} at a special value of the
Chern-Simons coupling only chiral boundary degrees of freedom exist. If
true, the theory could solve some of the problems with Witten's original
proposal~\cite{MW,m}.

However in ~\cite{GJ} a propagating mode was found even at the special, ``chiral'' point $\mu l =1$
(defined in eq.~(\ref{15}) below). This is not by itself in
contradiction with~\cite{SLS}, since the mode mode does not respect standard
Brown-Henneaux boundary conditions~\cite{BH}: near the boundary
it diverges linearly in the AdS radius. On the other hand, propagating
modes obeying Brown-Henneaux boundary conditions were found
in~\cite{DCWW,DCWW2}.\footnote{See also~\cite{DJJ,Car} for a canonical analysis of TMG, which
also shows a propagating degree of freedom at the chiral point.}
Those papers work in the Poincar\'e patch of AdS, 
which
only covers a part of the space; ref.~\cite{SLS} instead uses global
coordinates. This difference between coordinate systems makes direct
comparison of
the Poincar\'e patch modes with the global-coordinate ones difficult. Among
other things, the Poincar\'e patch energy does not coincide with
energy in global coordinates.
One is an element of the Lorentz subgroup of AdS$_3$ isometries:
$SO(1,2)\subset SO(2,2)$; the other is the (cover of) one $SO(2)$ in the
subgroup
$SO(2)\times SO(2) \subset SO(2,2)$. In the Poicar\'e patch the
global energy appears as the generator of dilatations.
Finally, the Poincar\'e patch
energy has a continuous spectrum while the global energy spectrum is discrete.

In this letter we will work in global coordinates and analyze the spectrum of
of topologically massive gravity at the chiral point using the same method as
ref.~\cite{SLS}. By applying appropriate generators of the AdS$_3$ isometry
group
$SO(2,2) \sim SL(2,R) \times \overline{SL(2,R)}$ to the linearly-divergent mode
of ref.~\cite{GJ}, we find modes that obey standard Brown-Henneaux boundary
conditions. They are all descendant of a field that is not quite primary: it
 transforms into a {  locally} 
pure gauge mode when hit by the $L_{+1},\bar{L}_{+1}$
generators of $SL(2,R) \times \overline{SL(2,R)}$. This gauge mode has zero
norm~\cite{SLS} { and it is pure gauge under diffeomorphism
that act on the boundary as left-moving conformal transformations.
After using this larger diffeomorphisms group to factor out the zero norm 
state}, the new modes fall into
a standard discrete representation of $SL(2,R) \times \overline{SL(2,R)}$. 
The representation is spanned by
$SL(2,R) \times \overline{SL(2,R)}$ descendants of a primary of weights $h=2,
\bar{h}=1$, and thus carries an intriguing---though as yet 
mysterious---kinship with the topologically
massive spin-one field that TMG reduces to at the chiral point~\cite{DCWW}. 

{  By factoring out all diffeomorphisms that act on the boundary as left 
moving conformal transformations one can define a theory where only 
the right moving boundary Virasoro algebra acts nontrivially on physical 
states~\cite{strom}. However this theory cannot be unitary, since the bulk state is
still not pure gauge and its norm is negative.}
\bigskip

The action of topological massive gravity is
\begin{equation}
S=+\frac{1}{2\pi }\int d^{3}x \, \sqrt{-g}\left( R+\frac{2}{l^{2}}\right) +\frac{%
1}{4\pi \mu }\int d^{3}x \, \varepsilon ^{\alpha \mu \nu }\left( \Gamma _{{
}\alpha \beta }^{\beta }\partial _{\mu }\Gamma _{{ }\nu \beta }^{\sigma
}+\frac{2}{3}\Gamma _{{ }\alpha \beta }^{\beta }\Gamma _{{ }\mu
\gamma }^{\sigma }\Gamma _{{ }\nu \beta }^{\gamma }\right) ,
\label{action}
\end{equation}%
where $\Lambda =-l^{-2}$ and $\mu $ is the coupling constant of the
Chern-Simons term. We choose a positive sign for the Einstein-Hilbert term.
With this choice, BTZ~\cite{BTZ} black holes have
positive energy for $\mu l > 1$, while massive gravitons have negative energy.

The equations of motion are
\begin{equation}
R_{\mu \nu }-\frac{1}{2}Rg_{\mu \nu }-\frac{1}{l^{2}}g_{\mu \nu }+\frac{1}{%
\mu }C_{\mu \nu }=0.
\end{equation}
The Cotton tensor $C_{\mu \nu }$ is defined as:
\begin{equation}
C_{{ }\nu }^{\mu }=\frac{1}{2}\varepsilon ^{\mu \alpha \beta }\nabla
_{\alpha }R_{\beta \nu }+\frac{1}{2}\varepsilon _{\nu }^{{ }\alpha
\beta }\nabla _{\alpha }R_{\beta }^{{ }\mu }.
\end{equation}

The Cotton tensor is the three-dimensional analog of the Weyl
tensor in the sense that $C_{\mu \nu }=0$ if and only if the metric is
conformally flat. The Cotton tensor vanishes for any solution to
Einstein gravity, so all GR solutions are also solutions of TMG.

In this note we want to determine whether TMG possesses degrees of freedom propagating
on an AdS background. In other words, we are interested in the perturbative
spectrum, i.e.~linearized fluctuations around empty AdS space. Thus we expand
$g_{\mu \nu }=\overline{g}_{\mu \nu
}+h_{\mu \nu }+\mathcal{O}(h^2)$, with $\overline{g}_{\mu \nu }$ the AdS$_{3}$
metric.

The perturbation must leave the metric asymptotically AdS$_{3}$. This
fixes the asymptotics to be~\cite{BH}:
\begin{eqnarray}\label{asymp}
g_{tt} &=&-r^{2}/l^{2}+\mathcal{O}\left( 1\right) {,}\qquad
g_{rr}=l^{2}/r^{2}+\mathcal{O}\left( r^{-4}\right) {,}\qquad g_{\phi
\phi }=r^{2}+\mathcal{O}\left( 1\right) \nonumber  \\
g_{r\phi } &=&\mathcal{O}\left( r^{-3}\right) ,\qquad g_{rt}=\mathcal{O}%
\left( r^{-3}\right) ,\qquad g_{t\phi }=\mathcal{O}\left( 1\right) .
\end{eqnarray}%
Here we have used a global coordinate system in which the AdS$_{3}$ metric is
\begin{equation}
ds^{2}=g_{\mu \nu }dx^{\mu }dx^{\nu }=-(1+r^{2}/l^{2})dt^{2}+\frac{dr^{2}}{%
(1+r^{2}/l^{2})}+r^{2}d\phi ^{2}.
\end{equation}%
$\phi $ is the angular direction, and the radial direction is $r\geq 0$.
The boundary is located at $r=\infty $.

It is convenient to write the AdS$_{3}$ metric in the
following form:
\bea
ds^{2}=l^{2}\left( -\cosh^{2}\rho \, dt^{2}+\sinh ^{2}\rho { }%
\, d\phi ^{2}+d\rho ^{2}\right).
\eea{adsm}
where we defined $r=l \sinh \rho$ and rescaled $t \to lt$. This coordinate system also covers the whole space, with the
boundary at $\rho =\infty$.
In these coordinates the asymptotics (\ref{asymp}) become
\bea
h_{\rho \rho} &\simeq &\mathcal{O}\left( e^{-2 \rho}\right) ,\qquad h_{\rho t}\simeq
\mathcal{O}\left( e^{-2 \rho}\right) ,\qquad h_{\rho \phi }\simeq \mathcal{O}\left(
e^{-2 \rho}\right) , \nonumber \\
h_{tt} &\simeq &\mathcal{O}\left( 1\right) ,\qquad h_{\phi \phi }\simeq
\mathcal{O}\left( 1\right) ,\qquad h_{t\phi }\simeq \mathcal{O}\left(
1\right) .
\eea{asymp2}

The isometry group of AdS$_{3}$ space is $SL(2,R)\times \overline{SL(2,R)}$,
and its generators are realized on scalar fields by%
\begin{eqnarray}
L_{0} &=&i\partial _{u},\qquad L_{\pm 1}=ie^{\pm iu}\left( \frac{\cosh 2\rho
}{\sinh 2\rho }\partial _{u}-\frac{1}{\sinh 2\rho }\partial _{v}\mp \frac{i}{%
2}\partial _{\rho }\right) , \\
\overline{L}_{0} &=&i\partial _{v},\qquad \overline{L}_{\pm 1}=ie^{\pm
iv}\left( \frac{\cosh 2\rho }{\sinh 2\rho }\partial _{v}-\frac{1}{\sinh
2\rho }\partial _{u}\mp \frac{i}{2}\partial _{\rho }\right).
\end{eqnarray}%
The two light-like variables introduced here are defined by
$u=t+\phi $ and $v=t-\phi $.

In the linear approximation about the solution $\overline{g}_{\mu \nu }$, the
graviton equations of motion take the form

\begin{equation}
\left( \mathcal{D}^{+}\mathcal{D}^{-}\mathcal{D}^{M}h\right) _{\mu \nu }=0
\label{12}
\end{equation}
where the metric is $g_{\mu \nu }=\overline{g}_{\mu \nu }+h_{\mu \nu }+%
\mathcal{O}(h^2)$, and where%
\begin{equation}
\left( \mathcal{D}^{\pm }{ }\right) _{\mu }^{{ }\nu }=\delta _{\mu
}^{{ }\nu }\mp l{ }\varepsilon _{\mu }^{{ }\alpha \nu }%
\overline{\nabla }_{\alpha },\qquad \left( \mathcal{D}^{M}{ }\right)
_{\mu }^{{ }\nu }=\delta _{\mu }^{{ }\nu }+\frac{1}{\mu }{ }%
\varepsilon _{\mu }^{{ }\alpha \nu }\overline{\nabla }_{\alpha }.
\end{equation}%
The covariant derivative $\overline{\nabla }_{\alpha }$ is defined
using the background metric $\overline{g}_{\mu \nu }$.

Since $\mathcal{D}^{-},$ $\mathcal{D}^{+}$ and $\mathcal{D}^{M}$
commute with each other, one can obtain all linearized solutions in terms of
three functions $h_{\mu \nu }^{\pm }$ and $h_{\mu \nu }^{M}$ which
obey\footnote{We defined $\psi _{\mu \nu }^{R}=\psi _{\mu \nu }^{+}$ and
$\psi _{\mu
\nu }^{L}=\psi _{\mu \nu }^{-}$.}
\beq
\left( \mathcal{D}^{+}h^{+}\right) _{\mu \nu }=\left( \mathcal{D}%
^{-}h^{-}\right) _{\mu \nu }=\left( \mathcal{D}^{M}h^{M}\right) _{\mu \nu }=0
\eeq{14}

We want to analyze the theory at the ``chiral'' point
\begin{equation}
\mu l=1,
\label{15}
\end{equation}
which could yield a chiral gravity, i.e. a theory in which only boundary modes of
definite chirality and black holes exist. According to ref.~\cite{SLS},
the theory at the chiral point is
consistent because all negative energy modes disappear. A first
observation is that, at the point $\mu l=1$, $\mathcal{D}%
^{M}=\mathcal{D}^{-}$.

In \cite{GJ} an additional solution---not considered
in~\cite{SLS}---was found. It is proportional to
$\partial_\mu h_{\mu\nu}|_{\mu=1/l}$,
which manifestly solves eq.~(\ref{12}). Its explicit form is
\begin{equation}
h_{\mu \nu }^{({new})}=\mbox{Re}\, \psi _{\mu \nu }^{({new})},\qquad
\mbox{with}\qquad \psi _{\mu \nu }^{({new})}=y(t,\rho ){ }\psi
_{\mu \nu }^{-}=e^{-2iu}{ }y(t,\rho ){ }H_{\mu \nu }(\rho ).
\end{equation}%
The function $y(t,\rho )$ is
\begin{equation}
y(t,\rho )=-\frac{i}{2}(u+v)-\log \left( \cosh \rho \right) ,
\end{equation}%
while the $H_{\mu \nu }(\rho )$ are the components of the tensor (in the $t, \phi,\rho$ basis)
\begin{equation}
H(\rho )=\left(
\begin{array}{ccc}
\tanh ^{2}\rho  & \tanh ^{2}\rho  & i\frac{\sinh \rho }{\cosh ^{3}\rho } \\
\tanh ^{2}\rho  & \tanh ^{2}\rho  & i\frac{\sinh \rho }{\cosh ^{3}\rho } \\
i\frac{\sinh \rho }{\cosh ^{3}\rho } & i\frac{\sinh \rho }{\cosh ^{3}\rho }
& -\frac{1}{\cosh ^{4}\rho }%
\end{array}%
{ }\right) .
\end{equation}

Now the crucial observation is that, while $\psi _{\mu \nu }^{({new})}$ and 
its descendants, of the form ${\cal L}_{L_{-1}}^n\psi _{\mu \nu }^{({new})}\,$ 
diverge linearly in $y(t,\rho )$ near the boundary,\footnote{That is, linearly
in $\rho$ and also linearly in time.} there also exist descendants that
satisfy the {  standard Brown-Henneaux} asymptotics.\footnote{For quasinormal modes in a
BTZ black hole background a similar fact has been independently
noticed in~\cite{S}.} 
{  Here we adopted the standard notation ${\cal L}_v$ for the Lie derivative along the vector
field $v$. From now on, to simplify notations, we shall denote by $L_n$, 
$\overline{L}_n$ both the vector field and the Lie derivative along it, whenever unambiguos.
Also, we must recall that metric fluctuations are not uniquely defined: two fluctuations 
that can be mapped into each other by diffeomorphisms that vanish at infinity as
\beq 
\zeta^\rho = {\cal O}(e^{-2\rho}), \qquad \zeta^t = {\cal O}(e^{-4\rho}), \qquad 
\zeta^\phi = {\cal O}(e^{-4\rho}),
\eeq{bhdiff}
represent the same physical state. We shall denote
equality up to these trivial diffeomorphisms by $\bh$. }

An example of one such state is obtained as follows: 
{  define first of all the tensor perturbation
\beq
Y_{\mu \nu }\equiv \overline{L}_{-1}\psi ^{({new})}_{\mu
\nu} = \frac{1}{2}e^{-iv}\tanh \rho { }\psi _{\mu \nu }^{-} + h_{\mu\nu} =
 \frac{1}{2} e^{-i(v+2u)}\tanh \rho { }H_{\mu \nu }(\rho ) + h_{\mu\nu} \label{Y} .
\end{equation}

A simple calculation shows that $h_{\rho\phi}={\cal O}[y(t,\rho)e^{-2\rho}]$, 
$h_{\rho t}={\cal O}[y(t,\rho)e^{-2\rho}]$, $h_{\rho\rho}={\cal O}[y(t,\rho)e^{-4\rho}]$;
therefore, $h_{\rho\phi}$, $h_{\rho t}$ do not obey the asymptotics~(\ref{asymp2}), 
while $h_{\rho\rho}$ does. Inspection of the AdS metric eq.~(\ref{adsm}) shows immediately that
$h_{\rho\phi}$ and $h_{\rho t}$ can be canceled up to terms with proper asymptotics by an 
infinitesimal diffeomorphism 
\beq
t\rightarrow t + \zeta^t, \qquad \phi \rightarrow \phi + \zeta^\phi, \qquad
\zeta^t,\zeta^\phi=\mbox{constant}\, e^{-i(v + 2u) -4\rho}[ y(t,\rho)+ {\cal O}(1)],
\eeq{diff}
which does not spoil the good asymptotics of any other component of the metric.

So, a bulk mode with proper boundary conditions is
\beq
X_{\mu\nu} = Y_{\mu\nu} + \nabla_\mu \zeta_\nu + \nabla_\nu \zeta_\mu = 
Y_{\mu\nu} + {\cal L}_\zeta \overline{g}_{\mu\nu}.
\eeq{X}
}

The field $X_{\mu\nu}$ defined above and its
$SL(2,R)\times \overline{SL(2,R)}$
descendants generate a tower of states with the standard Brown-Henneaux asymptotic behavior.

It is worth mentioning that, while the (2,0)-primary states corresponding to
solutions to $\left( \mathcal{D}^{-}\psi ^{-}\right) _{\mu \nu }=0$ satisfy
\begin{equation}
\left( L_{0}+\overline{L}_{0}\right) \psi ^{-}=2\psi ^{-},\qquad \left(
L_{0}-\overline{L}_{0}\right) =2\psi ^{-},
\end{equation}%
the state $\psi _{\mu \nu }^{({new})}$ is not primary, but it obeys
instead the following equation
\begin{equation}
\left( L_{0}+\overline{L}_{0}\right) \psi ^{({new})}=2\psi ^{({new}%
)}+\psi ^{-},\qquad \left( L_{0}-\overline{L}_{0}\right) \psi ^{({new}%
)}=2\psi ^{({new})}.
\end{equation}%
Since $L_{0}-\overline{L}_{0}=i\partial
_{\phi }$, is the angular momentum, this state carries one unit of angular
momentum: if it were
a primary, it would generate a spin-one representation of
$SL(2,R) \times \overline{SL(2,R)}$.

{  $X_{\mu\nu}$ too fails to be a conventional primary field because
$SL(2,R)\times \overline{SL(2,R)}$ descent operators do not annihilate it.
Instead:
\bea
L_{+1}X_{\mu\nu}&=& {\cal L}_{[L_{+1},\zeta]}\overline{g}_{\mu\nu},\qquad 
\overline{L}_{+1}X_{\mu\nu}=\psi ^{-}_{\mu\nu} + 
{\cal L}_{[\overline{L}_{+1},\zeta]}\overline{g}_{\mu\nu}, 
\nonumber \\
L_{0}X_{\mu\nu} &=& 2X_{\mu\nu}+ {1\over 2}\overline{L}_{-1}\psi^-_{\mu\nu} + 
{\cal L}_{[L_{0},\zeta]}\overline{g}_{\mu\nu},\qquad
\overline{L}_{0}X_{\mu\nu}=X_{\mu\nu}+ {1\over 2}\overline{L}_{-1}\psi^-_{\mu\nu} +
{\cal L}_{[\overline{L}_{0},\zeta]}\overline{g}_{\mu\nu}.\nonumber \\&&
\eea{21}
In writing these equations we have used standard properties of the Lie derivative, the 
commutation relations of $SL(2,R)\times \overline{SL(2,R)}$ and the equations
$L_0y=\overline{L}_0y=1/2$, $L_{+1}y=\overline{L}_{+1}y=0$.
We also exploited 
the fact that $\psi^-$ is a $(2,0)$-primary and that
$L_{+1},L_0,\overline{L}_{+1},\overline{L}_0$ are Killing vectors of the background AdS metric
$\overline{g}_{\mu\nu}$.

The asymptotic form of the vector field $\zeta$ defined in eq.~(\ref{diff}) is such that 
$[L_{+1},\zeta]$, $[\overline{L}_{+1},\zeta]$, $[L_{0},\zeta]$, 
$[\overline{L}_{0},\zeta]$ actually obey Brown-Henneaux boundary conditions and vanish at the 
AdS boundary; moreover, $\overline{L}_{-1}\psi^-=\overline{L}_{-1}L_{-2}
\overline{g}_{\mu\nu}\bh L_{-2}\overline{L}_{-1}\overline{g}_{\mu\nu}=0$.\footnote{We thank 
Alex Maloney for pointing this out to us.} The first equality follows from the definition of
$\psi^-$~\cite{SLS}, the second from Virasoro commutation relations, the third from 
$\overline{L}_{-1}$ being an isometry of the background metric. 
So eq.~(\ref{21}) can also be written as
\beq
L_{+1}X_{\mu\nu}\bh 0, \qquad \overline{L}_{+1}X_{\mu\nu} \bh \psi ^{-}_{\mu\nu}, \qquad 
L_{0}X_{\mu\nu} \bh  2X_{\mu\nu},\qquad \overline{L}_{0}X_{\mu\nu}\bh X_{\mu\nu}
\eeq{21a}
The second of these equations makes $X$ non-primary.}  Notice however that $\psi _{\mu \nu }^{-}$ is a pure gauge excitation. If
we define physical states modulo locally pure-gauge states,
$X_{\mu\nu}$ would be a true primary.

{  We can easily compute the norm of $X$ at the chiral point.  Start at a generic value of $\mu l$ and consider the mode $\bar L_{-1} \psi_M$, where $\psi_M$ is the massive graviton defined e.g in \cite{SLS}:  
\begin{equation}
 \bar L_{-1} \psi^M_{\mu \nu}   \bh  \bar L_{-1}(\psi^M - \psi^-)_{\mu \nu} + \nabla_\mu (\zeta^{\bar h}_\nu - \zeta^0_\nu) + \nabla_\nu  (\zeta^{\bar h}_\mu - \zeta^0_\mu) \equiv \bar h X^{\bar h}_{\mu \nu},
\end{equation}
where $\bar h =\mu l/2 - 1/2$ is the right-moving weight of the massive graviton and $\zeta_\mu^{\bar h} \equiv  e^{\bar h y} \zeta_\mu$ generates a trivial diffeomorphism.  The utility of this expression is that it converges pointwise to $X$ in the limit $\mu l \rightarrow 1$:  
\begin{equation}
X(\rho, u, v) = \lim_{\mu l \to 1} X^{\bar h}(\rho, u, v).
\end{equation}
We can now easily compute the norm of $X$:
\begin{equation}
\langle X | X \rangle = \lim_{\mu l \to 1} \langle X^{\bar h} | X^{\bar h} \rangle = \bar h^{-2} \langle \psi^M | \bar L_{-1}^\dagger \bar L_{-1} | \psi^M \rangle = 
2 \bar h^{-1} \langle \psi^M | \psi^M \rangle,
\end{equation}
where the second equality is true assuming the norm is invariant under Brown-Henneaux-trivial diffeomorphisms, and the last follows from the Virasoro algebra.  With the choice of sign which gives BTZ black holes positive energy, $ \langle \psi^M | \psi^M \rangle = \bar h C$ with $C$ negative~\cite{SLS}.  Hence the norm of $X$ is finite and negative.}

{ 
Metric fluctuations obeying Brown-Henneaux boundary conditions 
are completely determined by their asymptotically  non-vanishing components. 
In the case of $X_{\mu\nu}$, $\psi^-_{\mu\nu}$ and their
descendants, the only non-vanishing  component is $h_{uu}$. 
Therefore identification modulo the diffeomorphisms~(\ref{bhdiff})
tells us that a generic physical state takes the form
\bea
h_{uu}|_{\rho=\infty} &=& \sum_{n,m\geq 0} x_{m,n}e^{-(2+n)iu -(1+m)iv} + 
\sum_{n\geq 0}\psi^-_n e^{-(2+n)iu} + c.c. , \nonumber \\
h_{vv}|_{\rho=\infty} &=& \sum_{n\geq 0}\psi^+_n e^{-(2+n)iv} + c.c. \, .
\eea{bhphys}
Each of the Fourier coefficients $\psi_n^\pm, x_{n,m}$ is physical, so $X_{\mu\nu}$ is 
not a standard primary.  It is worth noting here that the asymptotics of this state at large $m,n$ are identical to the short wavelength asymptotics of the states found in \cite{DCWW}. 

At the chiral point $\mu l =1$, we can also  define a {\em different} theory, where states are identified modulo the larger 
group~\cite{strom}
\beq
\zeta^u = \epsilon(u) + {\cal O}(e^{-4\rho}),     \qquad
\zeta^v={1\over 2}e^{-2\rho} \partial_u^2 \epsilon(u) +  {\cal O}(e^{-4\rho}),   \qquad
\zeta^\rho= -{1\over 2}\partial_u \epsilon(u) + {\cal O}(e^{-2\rho}).
\eeq{stro}
This theory is chiral by construction. Physical states are defined by identifying those 
in~(\ref{bhphys}) modulo the Virasoro algebra generated on asymptotic states by eq.~(\ref{stro}). A possible gauge choice is to set
all $\psi^-_{n}=0$. This leaves yet unfixed the $SL(2,R)$ generated by $L_{\pm 1},L_0$.
We can use it to fix three real coefficients in $x_{m,n}$. {\em All other $x_{m,n}$ 
coefficients define distinct physical states}; by construction, they are
chiral primaries of the right-moving Virasoro algebra surviving factorization 
by~(\ref{stro}). 

}

In conclusion, {physical} states obeying the {\em standard} 
Brown-Henneax AdS$_{3}$ asymptotic
exist at the ``chiral" point $\mu l = 1$. They are descendants of an ``improper'' primary,
$\psi _{\mu \nu }^{({new})}$, which does not have the right asymptotics.
The lowest weight  state obeying the Brown-Henneaux asymptotics is
$X_{\mu\nu}$, given explicitly in eq.~(\ref{X}). 
It can be promoted to a true primary by  defining physical
states modulo { the Virasoro algebra~(\ref{stro}). Irrespective of the gauge group used to define physical states, the theory is non-unitary, because states with negative norm exist.}  
{ 
There is a strong case against topologically massive gravity 
being chiral {\em and unitary}
at the chiral point, which we summarize here:
\begin{itemize}
\item There is an extra mode $X$ at the chiral point  which obeys the Brown-Henneaux boundary conditions and which is not pure gauge (even with the prescription of \cite{strom}).
\item Modulo trivial diffeomorphisms $X$ can be obtained as a smooth limit of the $\bar L_{-1}$ descendent of the massive graviton in the limit $\mu l \rightarrow 1$. 
\item The asymptotic wavefunction for $X$ matches the Poincare-patch results of \cite{DCWW} at short wavelength.
\item Including $X$, the counting of states matches the canonical analyses of \cite{DJJ, Car, Park}.
\item The norm of $|X \rangle$ is negative.
\end{itemize}  }

A still unresolved and intriguing question  is how the
(2,1) representation we have found
relates to the fact that TMG at the chiral point can be thought of as a
topologically massive spin one field.
A better understanding the chiral spectrum of~\cite{DCWW}
may shed light on this connection.

A possible way out is if the theory at
the chiral point possesses another yet to be discovered gauge symmetry beyond diffeomorphisms (perhaps akin to the Weyl invariance of pure Chern-Simons gravity in flat space), which changes the definition of the energy and renders $X$ pure gauge.  Another possibility is that  our mode becomes non-normalizable at higher than leading order.\footnote{We thank A. Strominger for suggesting this possibility to us.}  However, this possibility appears to be in conflict with the non-perturbative canonical analyses of  \cite{DJJ, Car, Park}.

\subsection*{Acknowledgments}
The work of M.P. is supported in part by NSF grants PHY-0245068 and PHY-0758032, and that 
of M.K. by NSF CAREER grant PHY-0645435.  M.P. and M.K. would like to thank the Scuola 
Normale Superiore and the Galileo Galilei Institute in Florence for their kind hospitality
and support. G.G. is indebted to the members of Centro de Estudios
Cient\'{\i}ficos CECS, and to the members of the CCPP at New York University, for their hospitality.  We thank S. Deser, D. Grumiller, A. Maloney, and 
A. Strominger for fruitful e-mail exchanges.


\bigskip

\begin{thebibliography}{9}

\bibitem{SLS} W. Li, W. Song and A. Strominger, \textit{Chiral gravity in
three dimensions}, [arXiv:0801.4566].

\bibitem{DCWW}
  S.~Carlip, S.~Deser, A.~Waldron and D.~K.~Wise,
  \textit{Cosmological Topologically Massive Gravitons and Photons,'}
  Phys.\ Lett.\ B. {666} (2008) 272.


\bibitem{W} E.~Witten, \textit{Three-Dimensional Gravity Revisited},
  arXiv:0706.3359 [hep-th].

\bibitem{SLS2} W. Li, W. Song and A. Strominger, \textit{Comments on
"Cosmological Topological Massive Gravitons and Photons"}, [arXiv:0805.3101].

\bibitem{DJT1}
  S.~Deser, R.~Jackiw and S.~Templeton, \textit{Three-Dimensional Massive
Gauge Theories},
  Phys.\ Rev.\ Lett.\  {  48} (1982) 975.

\bibitem{DJT2}
  S.~Deser, R.~Jackiw and S.~Templeton, \textit{Topologically massive gauge
theories},
  Annals Phys.\  {  140}, 372 (1982)
  [Erratum-ibid.\  {  185}, 406.1988\ APNYA,281,409
(1988\ APNYA,281,409-449.2000)].

\bibitem{MW}
  A.~Maloney and E.~Witten, \textit{Quantum Gravity Partition Functions in
Three Dimensions},
  [arXiv:0712.0155].

\bibitem{m}
  J.~Manschot, \textit{AdS$_3$ Partition Functions Reconstructed},
  JHEP {  0710}, 103 (2007)
  [arXiv:0707.1159].

\bibitem{GJ} D. Grumiller and N. Johansson,\textit{\ Instability in
cosmological topological massive gravity at the chiral point},
[arXiv:0805.2610].

\bibitem{BH} J. Brown and M. Henneaux, \textit{Central charges in canonical
realization of asymptotic symmetries: An example from three-dimensional
gravity}, Comm. Math. Phys. \textbf{104} (1986) 207.

\bibitem{DCWW2} S. Carlip, S. Deser, A. Waldron and K. Wise, \textit{%
Topological massive AdS gravity}, [arXiv:0807.0486].

\bibitem{DJJ} D.~Grumiller, R.~Jackiw and N.~Johansson, \textit{Canonical analysis of 
cosmological topologically massive gravity at the chiral point},
  [arXiv:0806.4185].

\bibitem{Car}
  S.~Carlip, \textit{The Constraint Algebra of Topologically Massive AdS 
Gravity}, [arXiv:0807.4152].

\bibitem{strom} A.~Strominger, \textit{A Simple Proof of the Chiral Gravity Conjecture},
  [arXiv:0808.0506].

\bibitem{BTZ} M. Ba\~{n}ados, C. Teitelboim and J. Zanelli, \textit{The
black hole in three-dimensional space-time}, Phys. Rev. Lett. \textbf{69}
(1992) 1849, [arXiv:hep-th/9204099].

\bibitem{S} I. Sachs, \textit{Quasi-normal modes for logarithmic conformal
field theory}, [arXiv:0807.1844].

\bibitem{Park} M.~i.~Park, {\it Constraint Dynamics and Gravitons in Three Dimensions,} [arXiv:0805.4328].




\end{thebibliography}
\end{document}